\documentclass[preprintnumbers,amsmath,amssymb,twocolumn]{revtex4}
\usepackage{graphicx}
\usepackage{dcolumn}
\usepackage{bm}
\usepackage{epsfig}

\begin{document}
\title{Superluminal light propagation via quantum interference in decay channels}
\author{R. Arun}
\affiliation{Department of Physics, School of Basic $\&$ Applied Sciences, Central University of
Tamilnadu, Thiruvarur 610101, Tamilnadu, India.}
\email{rarun@cutn.ac.in}

\begin{abstract}
We examine the propagation of a weak probe light through a coherently driven $Y$-type system. Under
the condition that the excited atomic levels decay via same vacuum modes, the effects of
quantum interference in decay channels are considered. It is found that the interference in decay channels results in
a lossless anomalous dispersion between two gain peaks. We demonstrate that the probe pulse propagation can in principle
be switched from subluminal to superluminal due to the decay-induced interference. We also show that the system
exhibits a high index of refraction with negligible absorption for the driving fields. A dressed-state picture
of the atom-light interaction is described to explain the numerical results.
\end{abstract}

\maketitle
\newpage
\section{Introduction}
\vspace{-1.5em}
The study of subluminal and superluminal light propagation has been receiving a great deal of attention during the last
couple of decades. This is mainly due to a number of impressive experiments that demonstrated both slow \cite{kasapi,meschede,hau,kash,
budker,turukhin,bigelow,lukin2} and fast light propagation \cite{wang} in dispersive media. Most studies of light propagation employ the
standard electromagnetically induced transparency (EIT) setup to control the group velocity of light pulses \cite{kasapi,meschede,hau,kash,
budker,turukhin}. In these schemes, the absorption of light is very low while the dispersion has a steep positive slope leading to low values
of the group velocity. The reduction of group velocity $(v_g < c)$ to less than the vacuum speed of light (subluminal light propagation) has been
demonstrated both in atomic vapor \cite{kasapi,meschede,hau,kash,budker} and solid state materials \cite{turukhin,bigelow}. In the extreme case of
ultraslow light propagation, light pulses can also be stopped $(v_g = 0)$ \cite{lukin1,scully1,gsa1} and stored as demonstrated in many experiments
\cite{lukin2}. On the other hand, a light pulse can also travel with a group velocity exceeding the vacuum speed of light $(v_g > c)$ or even negative
$(v_g < 0)$. Such propagation of light pulses, termed as superluminal light propagation, occurs in anomalous dispersion media. It was pointed out
that superluminal light propagation is not in conflict with causality or special relativity \cite{review} but a consequence of wave interference
phenomena \cite{mborn}. The early work of Steinberg and Chiao suggested that an anomalous dispersion having a steep negative slope can appear between
two closely spaced gain lines \cite{steinberg}. Based on the work of Steinberg and Chiao \cite{steinberg}, Wang {\it et al.} reported the first
experimental demonstration of superluminal pulse propagation using the transparent anomalous dispersion in a Raman gain doublet \cite{wang}. There are
also parallel developments in achieving both subluminal and superluminal light in atomic vapor \cite{godone} and solid systems \cite{solids}. In
addition, some theoretical works showed the possibility of switching light propagation from subluminal to superluminal \cite{gsa2,mahmoudi,joshi,wilson}.
Agarwal {\it et al.} demonstrated that the dispersion in $\Lambda$ systems can be manipulated by using an additional field coupling the metastable
ground states \cite{gsa2}. Mahmoudi {\it et al.} explained how the use of incoherent pump fields can influence the dispersion characteristics of the
system \cite{mahmoudi}. The group velocity control has been shown to be possible due to interaction of the atomic system with squeezed vacuum
reservoirs \cite{joshi}. Further, Bortman-Arbiv {\it et al.} reported that the relative phase between driving fields can control the dispersion in
$V$ systems \cite{wilson}.

In this paper, we exploit the fact that decay processes can alter the absorption/dispersion of light propagation in a medium as a result of quantum
interference in spontaneous decay channels. The interference occurs when two or more atomic transitions sharing common vacuum modes lead to a
spontaneous emission of identical photons. An important result of such decay-induced interference is the generation of a coherence between the
excited atomic states that are decaying to the common ground state \cite{gsa3}. A variety of interesting features arising from the coupling of atomic
transitions due to decay-induced interference has been reported
\cite{wilson,gsa3,quench,zhu1,zhu2,swain,gsa4,autler,macovei,carreno,hou,arun1,kerr,soliton,arun2}.
Examples include phase control of group velocity \cite{wilson}, population trapping \cite{gsa3}, fluorescence quenching \cite{quench},
dark lines in emission spectrum \cite{zhu1}, spontaneous emission cancellation \cite{zhu2}, ultranarrow spectral lines in fluorescence \cite{swain},
delayed formation of trapping state \cite{gsa4}, gain in Autler-Townes doublet \cite{autler}, phase control of collective dynamics \cite{macovei},
control of group velocity with squeezed reservoir \cite{carreno}, enhancement of two-photon absorption \cite{hou}, splitting of resonances \cite{arun1},
enhancement of self-Kerr nonlinearity \cite{kerr}, soliton formation \cite{soliton}, and enhanced spectral squeezing in fluorescence \cite{arun2}, etc.

Our model system for realizing quantum interference in spontaneous emissions is shown schematically in Fig. 1. This model has been extensively investigated
in different contexts \cite{hou,arun1,kerr,soliton,arun2}. The atom has a Y-type configuration in which two near-degenerate excited states decay via
common vacuum modes to the middle state. The atom in the middle state can further undergo spontaneous transitions to the ground state. The cascade decays
$|1\rangle \rightarrow |3\rangle \rightarrow |4\rangle$ and $|2\rangle \rightarrow |3\rangle \rightarrow |4\rangle$ of the excited atom give rise to
an emission of identical pair of photons and hence quantum interference exists in emission pathways. We investigate the role of the quantum interference in
decay channels on the pump-probe spectroscopy of the system. We show that the quantum interference manifests itself as gain doublet in the absorption
spectrum with an anomalous dispersion between the gain peaks.

\begin{figure}[t]
\hspace{-4em}
    \includegraphics[width=8cm]{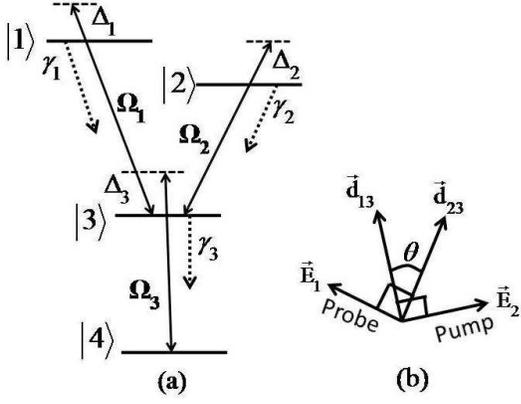}
    \caption{(a) The level scheme of a $Y$-type four-level atom driven by pump and probe fields. (b) The arrangement of field polarisations with respect
to the atomic dipole moments. The pump and probe fields each act on only one transition.}
\vspace{-1.5em}
\end{figure}
The paper is arranged as follows. In Sec. II, we present the atomic model and its basic dynamical equations describing the interaction of the atom
with probe and pump fields. The effects of decay-induced interference on the absorption/dispersion of the probe field are then discussed in Sec. III.
We show the possibility of gain features in the probe absorption spectrum as well as switching from normal to anomalous behavior in the probe
dispersion profile. In Sec. IV, we develop an analysis using dressed states which explains the gain features in the probe absorption spectrum.
Section V is devoted to discuss the possibility of population trapping in the dressed states and its effect on the dispersion of the pump fields.
\vspace{-1em}
\section{atomic model and basic equations}
\vspace{-1em}
A four-level atom in the $Y$-type configuration is shown schematically in Fig. 1(a). The atom has two near-degenerate excited states $|1\rangle$ and $|2\rangle$
which decay spontaneously via same vacuum modes to the middle state $|3\rangle$ with rates $2 \gamma_1$ and $2 \gamma_2$, respectively. A spontaneous emission
from the state $|3\rangle$ (decay rate $2 \gamma_3$) leaves the atom in its ground state $|4\rangle$. It is assumed that direct transitions between the excited
and ground states $(|1\rangle,|2\rangle\leftrightarrow|4\rangle)$ as well as between the excited states $(|1\rangle\leftrightarrow|2\rangle)$ are dipole
forbidden. We consider a situation in which pump and probe fields act on different transitions in the atom. The transition $|2\rangle\leftrightarrow|3\rangle$
of frequency $\omega_{23}$ is driven by a pump field (amplitude $E_2$, phase $\phi_2$) of Rabi frequency $2 \Omega_2=2\vec{d}_{23}.\vec{E_2}/\hbar$. The
transition $|3\rangle\leftrightarrow|4\rangle$ of frequency $\omega_{34}$ is driven by a pump field (amplitude $E_3$, phase $\phi_3$) of Rabi frequency
$2 \Omega_3 = 2 \vec{d}_{34}.\vec{E_3}/\hbar$. A probe field (amplitude $E_1$, phase $\phi_1$) of Rabi frequency $2 \Omega_1 = 2 \vec{d}_{13}.\vec{E_1}/\hbar$
is set to act on the transition $|1\rangle\leftrightarrow|3\rangle$ of frequency $\omega_{13}$. Here $\vec{d}_{ij}$ is the electric transition dipole moments
(assumed to be real) between states $|i\rangle$ and $|j\rangle$. We assume that the transition dipole moments $\vec{d}_{13}$ and $\vec{d}_{23}$ are at
an angle $\theta$, $\theta \neq 0$, for the pump $(\vec{d}_{13}.\vec{E_2}=0)$ and probe $(\vec{d}_{23}.\vec{E_1}=0)$ fields to act on only one transition as
shown in Fig. 1(b). The Hamiltonian for this system in the dipole and rotating wave approximations will be
\begin{flalign}
H = &~\hbar \omega_{14} A_{11} + \hbar \omega_{24} A_{22} + \hbar \omega_{34} A_{33}
&\nonumber \\
& - \hbar (\Omega_1 A_{13} e^{-i (\omega_1 t - \phi_1)} + \Omega_2 A_{23} e^{-i (\omega_2 t - \phi_2)} + \hbox{H.c.}) &
\nonumber \\
& - \hbar(\Omega_3 A_{34} e^{-i (\omega_3 t - \phi_3)} + \hbox{H.c.}),&
\end{flalign}
where the zero of energy is defined at the ground state $|4\rangle$ and $\hbar\omega_{mn}$ is the energy difference between the states $|m\rangle$ and
$|n\rangle$. The operators $A_{mn}=|m\rangle\langle n|$ represents the atomic population operators for $m=n$ and transition operators $m\neq n$.

We use the master equation framework to study the dynamics of the system. In the Schr\"{o}dinger picture, the evolution of the density operator $\tilde{\rho}$
of the atom obeys
\begin{equation}
\frac{d\tilde{\rho}}{dt} = \frac{1}{i\hbar} [H,\tilde{\rho}] + L\tilde{\rho},
\end{equation}
where the Liouvillian operator $L\tilde{\rho}$ describes the damping terms due to spontaneous emission. It is convenient to use an interaction picture
defined by the unitary transformation
\begin{eqnarray}
U &=& \exp\left\{[i(\omega_2 t - \phi_2) + i(\omega_3 t - \phi_3)]A_{11} \right. \nonumber \\
&&~~~~~~~ + [i(\omega_2 t - \phi_2) + i(\omega_3 t - \phi_3)]A_{22} \nonumber \\
&&~~~~~~~ \left. + [i(\omega_3 t - \phi_3)]A_{33} \right\}. \nonumber
\end{eqnarray}
In the interaction picture, the master equation for the reduced density operator $\rho = U \tilde{\rho} U^{\dag}$ of the atom takes the form
\begin{equation}
\frac{d\rho}{dt} = \frac{1}{i\hbar} [H_I,\rho] + L\rho, \label{mastereq}
\end{equation}
where the Hamiltonian in the interaction picture is
\begin{flalign}
H_I = &~\hbar (W_{12} - \Delta_2 - \Delta_3) A_{11} - \hbar (\Delta_2 + \Delta_3) A_{22} &  \nonumber \\
& - \hbar \Delta_3 A_{33} -\hbar (\Omega_1 A_{13} e^{-i(\delta t - \Phi)} + \Omega_1 A_{31} e^{i(\delta t - \Phi)}) &\nonumber \\
& - \hbar (\Omega_2 A_{23} + \Omega_2 A_{32}) - \hbar(\Omega_3 A_{34} + \Omega_3 A_{43}), & \label{ham}
\end{flalign}
and the damping term is
\begin{eqnarray}
L\rho &=& -\gamma_1 (A_{11} \rho - 2 A_{31} \rho A_{13} + \rho A_{11}) \nonumber \\
&& -\gamma_2 (A_{22} \rho - 2 A_{32} \rho A_{23} + \rho A_{22}) \nonumber \\
&& -\gamma_{12} (A_{21} \rho - 2 A_{31} \rho A_{23} + \rho A_{21}) \nonumber \\
&& -\gamma_{12} (A_{12} \rho - 2 A_{32} \rho A_{13} + \rho A_{12}) \nonumber \\
&& -\gamma_3 (A_{33} \rho - 2 A_{43} \rho A_{34} + \rho A_{33}). \label{dterm}
\end{eqnarray}
In Eq. (\ref{ham}), $\Delta_2=\omega_2-\omega_{23}$ $(\Delta_3=\omega_3-\omega_{34})$ corresponds
to the detuning of the field acting on $|2\rangle\leftrightarrow|3\rangle$ $(|3\rangle\leftrightarrow|4\rangle)$
transition, $\delta=\omega_1-\omega_2$ $(\Phi=\phi_1-\phi_2)$ gives the frequency (phase) difference of
the probe and pump fields, and $\hbar W_{12}$ is the energy separation between the excited atomic states. The cross-damping
terms $\gamma_{12}$ in Eq. (\ref{dterm}) arise from the quantum interference in decay channels $|1\rangle
\rightarrow |3\rangle$ and $|2\rangle\rightarrow|3\rangle$. Here $\gamma_{12}=\sqrt{\gamma_1\gamma_2}\cos(\theta)$
is referred to as interference parameter, which depends on the alignment $(\theta)$ of the transition dipole moments
$\vec{d}_{13}$ and $\vec{d}_{23}$.

The master equation (\ref{mastereq}), when written in the atomic state basis, yields the following equations for the
density matrix elements:
\begin{flalign}
\dot{\rho}_{11} =&-2\gamma_1 \rho_{11} - \gamma_{12} (\rho_{12} + \rho_{21}) + i \Omega_1 e^{-i(\delta t - \Phi)} \rho_{31}
&\nonumber \\
&-i\Omega_1 e^{i(\delta t - \Phi)} \rho_{13},&
\label{rho1}
\end{flalign}
\begin{flalign}
\dot{\rho}_{22} =&-2\gamma_2 \rho_{22} - \gamma_{12} (\rho_{12} + \rho_{21}) +
i \Omega_2 (\rho_{32} - \rho_{23}),&
\end{flalign}
\begin{flalign}
\dot{\rho}_{33} =&~2\gamma_1 \rho_{11} + 2 \gamma_2 \rho_{22} - 2 \gamma_3 \rho_{33}
+ 2 \gamma_{12} (\rho_{12} + \rho_{21})&  \nonumber \\
&+ i \Omega_1 e^{i(\delta t - \Phi)} \rho_{13} - i \Omega_1 e^{-i(\delta t - \Phi)} \rho_{31}& \nonumber  \\
&+ i \Omega_2 (\rho_{23} - \rho_{32}) + i \Omega_3 (\rho_{43} - \rho_{34}),&
\end{flalign}
\begin{flalign}
\dot{\rho}_{12}=&- (\gamma_1 + \gamma_2 + i W_{12}) \rho_{12} - \gamma_{12} (\rho_{11} + \rho_{22})& \nonumber \\
& + i \Omega_1 e^{-i(\delta t - \Phi)} \rho_{32} - i \Omega_2 \rho_{13},&
\end{flalign}
\begin{flalign}
\dot{\rho}_{13} =&-[\gamma_1 + \gamma_3 + i (W_{12}-\Delta_2)] \rho_{13} - \gamma_{12} \rho_{23}& \\
& + i \Omega_1 e^{-i(\delta t - \Phi)} (\rho_{33} - \rho_{11}) - i \Omega_2 \rho_{12} - i \Omega_3 \rho_{14},&
\nonumber
\end{flalign}
\begin{flalign}
\dot{\rho}_{23} =&-(\gamma_2 + \gamma_3 - i \Delta_2) \rho_{23} - \gamma_{12} \rho_{13}
- i \Omega_1 e^{-i(\delta t - \Phi)} \rho_{21}& \nonumber \\
& + i \Omega_2 (\rho_{33} - \rho_{22}) - i \Omega_3 \rho_{24},&
\end{flalign}
\begin{flalign}
\dot{\rho}_{34} =&-(\gamma_3 - i \Delta_3) \rho_{34}  + i \Omega_1 e^{i(\delta t - \Phi)} \rho_{14}
+ i \Omega_2 \rho_{24}& \nonumber \\
& + i \Omega_3 (\rho_{44} - \rho_{33}),&
\end{flalign}
\begin{flalign}
\dot{\rho}_{14} =&-[\gamma_1 + i (W_{12} - \Delta_2 - \Delta_3)] \rho_{14} - \gamma_{12} \rho_{24}
\nonumber & \\
& + i \Omega_1 e^{-i(\delta t - \Phi)} \rho_{34} - i \Omega_3 \rho_{13},&
\end{flalign}
\begin{flalign}
\dot{\rho}_{24} =&-[\gamma_2 - i (\Delta_2 + \Delta_3)] \rho_{24} - \gamma_{12} \rho_{14}
+ i \Omega_2 \rho_{34}& \nonumber \\
&  - i \Omega_3 \rho_{23}.&
\label{rho9}
\end{flalign}
As seen in Eqs. (\ref{rho1})-(\ref{rho9}), the excited state populations $(\rho_{11},\rho_{22})$ and the coherences
$(\rho_{12},\rho_{21})$ are coupled through the interference parameter $\gamma_{12}$. Physically, this corresponds to the
fact that the population can be transferred between the excited atomic states due to spontaneous emission from these states.
The spontaneous emission along one transition can drive its neighboring transition because of the common vacuum modes.
If the dipole moments $(\vec{d}_{13},\vec{d}_{23})$ are nearly parallel $(\theta\approx0)$, then $\gamma_{12}\approx
\sqrt{\gamma_1\gamma_2}$ and the interference effects are maximal, whereas there is no decay-induced interference if the
dipole moments are perpendicular $(\theta=90^{\circ})$ with $\gamma_{12}=0$.

\section{calculation of the susceptibility and group velocity}
\vspace{-1em}
In the following we shall calculate the susceptibility for the probe field in the linear regime
assuming that the probe field is weak enough $(\Omega_2,\Omega_3\gg \Omega_1 \ll \gamma_1,\gamma_2,\gamma_3)$.
The susceptibility of the medium depends on the atomic response to the probe field through the coherence term
$(\rho_{13})$ of the density matrix. To calculate the density matrix elements, we rewrite the equations (\ref{rho1})-(\ref{rho9})
in a compact form as
\begin{equation}
\frac{d}{dt} \hat{R} + \Sigma = M \hat{R},\label{matrix}
\end{equation}
where $\hat{R}$ is a column vector containing the density matrix elements
\begin{eqnarray}
\hat{R}=&\left(\rho_{11},\rho_{22},\rho_{33},\rho_{12},\rho_{13},\rho_{23},\rho_{14},\rho_{24}, \right.& \nonumber \\
& \left. \times \rho_{34},\rho_{21},\rho_{31},\rho_{32},\rho_{41},\rho_{42},\rho_{43}\right)^{T},&
\label{psidef}
\end{eqnarray}
and the inhomogeneous term $\Sigma$ is also a column vector with non-zero components $\Sigma_9=-i\Omega_3$ and $\Sigma_{15}=i\Omega_3$.
This term arises because we have eliminated the ground-state population $\rho_{44}$ by using the trace condition $\rho_{11}+\rho_{22} +\rho_{33}
+\rho_{44}=1$. In Eq. (\ref{matrix}), $M$ is a $15\times15$ matrix independent of the density matrix elements and contains time-independent
as well as time-dependent terms. The components of $M$ can be obtained explicitly using Eqs. (\ref{rho1})-(\ref{rho9}) and separated into terms
with different time dependencies,
\begin{equation}
M = M_0 + \Omega_1 M_1 e^{-i(\delta t - \Phi)} + \Omega_1 M_{-1} e^{i(\delta t - \Phi)}, \label{fmat}
\end{equation}
where the matrices $M_0$ and $M_{\pm1}$ have time-independent elements. Since the time dependence of $M$ is periodic, the solution to Eq.(\ref{matrix})
can be obtained using Floquet theorem. The solution $\hat{R}$ can be expanded into terms oscillating at harmonics of the detuning $\delta$:
\begin{equation}
\hat{R} = \hat{R}^{0} + \Omega_1 \hat{R}^{+} e^{-i(\delta t - \Phi)} + \Omega_1 \hat{R}^{-} e^{i(\delta t - \Phi)} + O(\Omega_1^2) + ...
\label{floquet}
\end{equation}
In the weak probe field approximation, the terms of order $\Omega_1^2$ or more will be neglected in the Floquet expansion (\ref{floquet}). Thus, on
combining Eqs. (\ref{matrix})-(\ref{floquet}), the steady-state solutions for $\hat{R}^0$ and $\hat{R}^{\pm}$ can be obtained as \cite{evers}
\begin{eqnarray}
\hat{R}^{0} &=& M_0^{-1} \Sigma, \nonumber \\
\hat{R}^{+} &=& - (M_0 + i \delta)^{-1}M_1 \hat{R}^{0},  \label{matsol}\\
\hat{R}^{-} &=& - (M_0 - i \delta)^{-1} M_{-1} \hat{R}^{0}. \nonumber
\end{eqnarray}

The susceptibility of the medium is related to the 5th component (the coherence $\rho_{13}$) of $\hat{R}$ which oscillates in phase with the probe field.
It is given by \cite{qbook}
\begin{equation}
\chi = \frac{N |\vec{d}_{13}|^{2}}{\epsilon_o \hbar} [\hat{R}^{+}]_5, \label{suscept}
\end{equation}
where $N$ is the number density of atoms in the medium. The real [Re$(\chi)$] and imaginary [Im$(\chi)$] parts of the susceptibility corresponds to the dispersion
and absorption of the probe beam, respectively. In our notation, if Im$(\chi)>0$, the probe field is absorbed, whereas Im$(\chi)<0$ corresponds to the probe
gain (amplification). Note that the steady-state solutions of the density matrix elements (\ref{matsol}) and the susceptibility (\ref{suscept}) are
independent of the relative phase $(\Phi)$ of the probe and pump fields. This feature is expected for a weak probe field consistent with the results in earlier
publications \cite{gsa1,gsa4,evers}. The susceptibility (\ref{suscept}) gives the absorption/dispersion characteristics of a continuous-wave (CW) probe light.
However, if the probe field is a pulse acting on the transition $|1\rangle\leftrightarrow|3\rangle$, the relevant quantity of interest is the group velocity
of the pulse. The group velocity of the probe pulse is related to the susceptibility (\ref{suscept}) for the $|1\rangle\leftrightarrow|3\rangle$
transition:
\begin{figure}[t]
\centering
\vspace{-0.4em}
   \includegraphics[width=8cm]{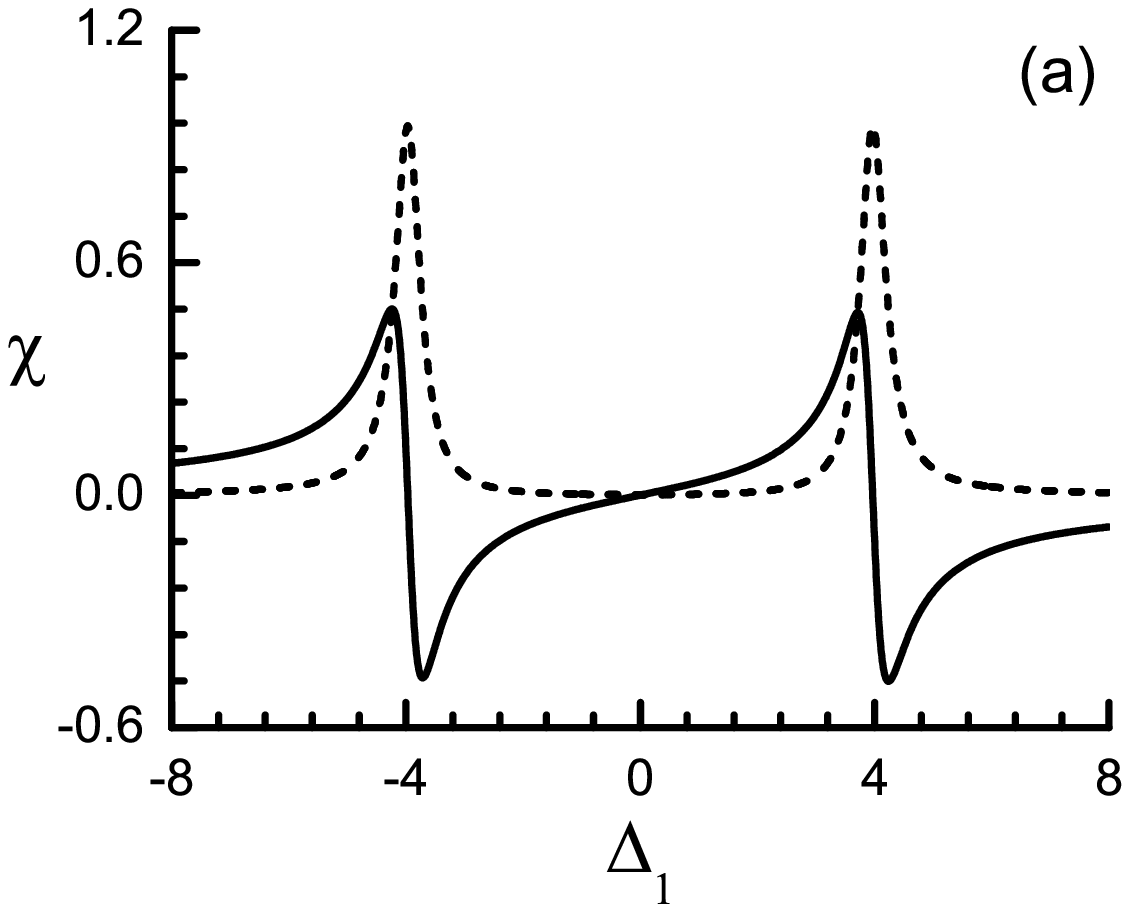} \\
   \includegraphics[width=8cm]{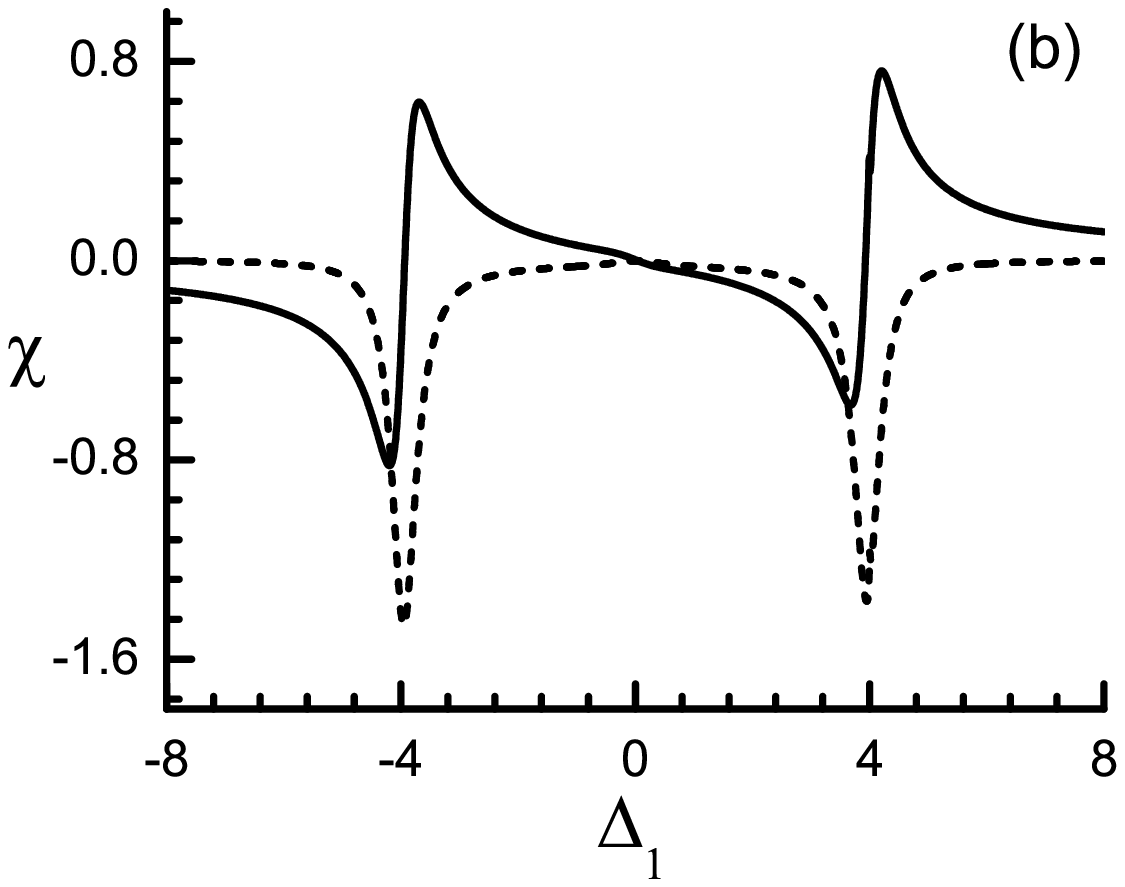}
   \vspace{-1.5em}
   \caption{Real (solid curve) and imaginary (dashed curve) parts of the susceptibility
   as a function of the probe detuning $\Delta_1$ for the parameters $\gamma_2 = 1$,
   $\gamma_1 = \gamma_3 = 0.01$, $\Delta_2 = \Delta_3 = 0$, $\Omega_2 = \Omega_3 = 4/\sqrt{2}$,
   $W_{12} = -\sqrt{\Omega_2^2 +\Omega_3^2}$, and (a) $\theta = 90^{\circ}$ and (b) $\theta = 15^{\circ}$.}
   \vspace{-1em}
\end{figure}
\begin{equation}
v_g = \frac{c}{1 + (\omega_1/2) (\partial\hbox{Re}\chi(\omega_1)/\partial\omega_1)}. \label{gvel}
\end{equation}
Clearly, the slope of the dispersion $(\partial\hbox{Re}\chi(\omega_1)/\partial\omega_1)$ at the central frequency $\omega_1$ of the pulse decides
the nature of pulse propagation. When the slope of the dispersion is positive, then $v_g<c$, indicating subluminal pulse propagation. On the other hand,
if the slope of the dispersion is negative, then $v_g>c$ or $v_g<0$ and the pulse propagation is in the superluminal region.

In order to study the probe absorption and dispersion features, we obtain the steady-state solution of Eq. (\ref{matrix}) numerically using
Eq. (\ref{matsol}). Both the presence $(\gamma_{12}\neq0)$ and absence $(\gamma_{12}=0)$ of quantum interference in decay channels are considered in the
numerical calculations. In Fig. 2, the susceptibility is plotted as a function of the probe detuning $\Delta_1 = \omega_1-\omega_{13}$, which is related to
the probe-pump detuning $(\delta)$ by $\Delta_1=\Delta_2+\delta-W_{12}$. We use dimensionless quantities by scaling all the frequency parameters such as
decay rates, detunings, and Rabi frequencies in units of $\gamma_2$. The susceptibility is scaled in units of $N|\vec{d}_{13}|^{2}/\epsilon_o\hbar\gamma_2$.
As seen in Fig. 2(a), the dashed curve (absorption spectrum) shows the usual Autler-Townes absorption components at $\Delta_1=\pm\sqrt{\Omega_2^2+\Omega_3^2}$
in the absence of interference parameter $(\gamma_{12}=0$). The dispersion curve (solid curve) in Fig. 2(a) shows a positive slope (normal dispersion) in the
region between the absorption peaks which implies subluminal pulse propagation for $\gamma_{12}=0$. The situation changes significantly when the interference
\begin{figure}[t]
\centering
   \includegraphics[width=8cm]{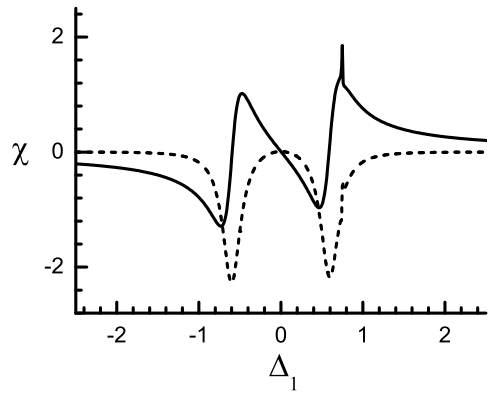}
   \vspace{-1.5em}
   \caption{Real (solid curve) and imaginary (dashed curve) parts of the susceptibility
   as a function of the probe detuning $\Delta_1$ for the parameters $\gamma_2 = 1$,
   $\gamma_1 = \gamma_3 = 0.01$, $\Delta_2 = \Delta_3 = 0$, $\Omega_2 = \Omega_3 = 0.75/\sqrt{2}$,
   $W_{12} = -\sqrt{\Omega_2^2 +\Omega_3^2}$, and $\theta = 15^{\circ}$.}
   \vspace{-1.5em}
\end{figure}
parameter $(\gamma_{12}\neq 0$) is included in the analysis. In Fig. 2(b), we show the probe-field line profiles for the same parameters of Fig. 2(a)
with $\gamma_{12} = 0.97\sqrt{\gamma_1\gamma_2}$ . The dashed curve (absorption spectrum) in Fig. 2(b) shows the remarkable result that the quantum
interference in decay channels gives rise to gain profiles instead of the absorption features. For zero detunings $(\Delta_2=\Delta_3=0)$ of the
pump fields, the gain doublet appears only for a parameter choice $W_{12}=\pm\sqrt{\Omega_2^2+\Omega_3^2}$ and $\gamma_1,\gamma_3\ll\gamma_2$.
These results are quite similar to those reported for the pump-probe spectroscopy in $V$ systems \cite{autler}. In contrast, a new feature arises in
the probe dispersion as shown by the solid curve in Fig. 2(b). The slope of the dispersion is seen  to be negative (anomalous dispersion) in the
region between the gain peaks, indicating superluminal light propagation. Thus, one finds that the effect of decay-induced interference is to switch
the light propagation from subluminal to superluminal domain.

Note that the separation $2 |W_{12}| = 2\sqrt{\Omega_2^2+\Omega_3^2}$ between the gain peaks depend on the Rabi frequencies $\Omega_2, \Omega_3$ of
the pump fields.  For a small splitting $W_{12}= -0.75 \gamma_2$ of the excited doublet as in Fig. 3, the gain peaks are closely spaced and the
dispersion can have a large negative slope exhibiting strong superluminal effects. To show the superluminal effect in a quantitative way, we rewrite
the group velocity $(\ref{gvel})$ using Eq. (\ref{suscept}) as
\vspace{-1.4em}
\begin{equation}
\frac{c}{v_g} = 1 + K \frac{\partial\hbox{Re}(\gamma_2 [\hat{R}^{+}]_5)}{\partial (\gamma_2^{-1} \omega_1)~~~~}, \label{vg}
\end{equation}
where
\vspace{-1em}
\begin{equation}
K = \frac{\omega_1 N |\vec{d}_{13}|^{2}}{2 \epsilon_o \hbar \gamma_2^2} = \frac{3 \pi N c^3 \gamma_1}{\omega_1^2 \gamma_2^2},
\label{kval}
\end{equation}
\begin{figure}[t]
\hspace{-3.5em}
   \includegraphics[width=7cm]{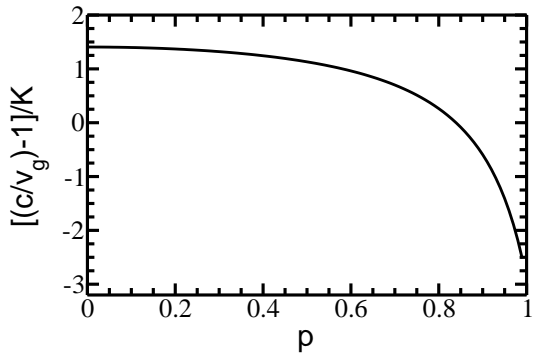}
   \vspace{-1em}
   \caption{Normalized group velocity $[(c/v_g) - 1]/K$ at $\Delta_1 = 0$ as a function of the interference parameter
   $p = \cos(\theta) = \gamma_{12}/\sqrt{\gamma_1 \gamma_2}$. The other parameters for the calculation are the same as in Fig. 3.}
   \vspace{-1em}
\end{figure}
is a dimensionless quantity. In Eq. (\ref{kval}), we have used, $2 \gamma_1=|\vec{d}_{13}|^{2}\omega_1^3/3\pi\epsilon_o\hbar c^3$,
the spontaneous decay rate of the $|1\rangle \leftrightarrow |3\rangle$ transition. An inspection of Eq. (\ref{vg}) reveals that
the slope of the dispersion is equal to $(c/v_g) - 1$ in units of $K$. Figure 4 displays the normalized group velocity $[(c/v_g) - 1]/K$,
calculated at the line centre $(\Delta_1 = 0)$, as a function of the interference parameter $p\equiv\gamma_{12}/\sqrt{\gamma_1\gamma_2}$.
The graph shows that the slope of the dispersion decreases monotonically with increasing $\gamma_{12}$ and becomes negative after
$\gamma_{12} \approx 0.8 \sqrt{\gamma_1\gamma_2}$. The maximum negative slope occurs near $\gamma_{12} = \sqrt{\gamma_1 \gamma_2}$
resulting in a large superluminality.
\begin{figure}[t]
   \includegraphics[width=9cm]{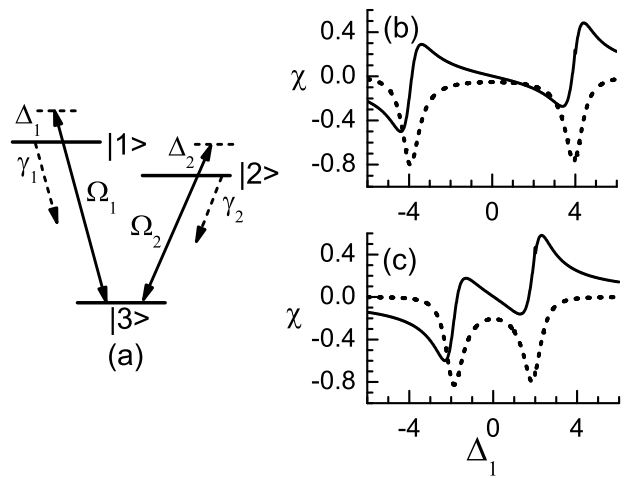}
\vspace{-2em}
   \caption{(a) The level scheme of a $V$-type three-level atom interacting with pump and probe fields. (b) Susceptibility $(\chi)$ versus
   the probe detuning $\Delta_1$ of the $V$-type atom for $\Omega_2 = 4 \gamma_2$. (c) Susceptibility $(\chi)$ versus the probe detuning
   $\Delta_1$ of the $V$-type atom for $\Omega_2 = 2 \gamma_2$. The solid (dashed) curve represents the real (imaginary) part of the susceptibility.
   The common parameters for (b) and (c) are $\gamma_2 = 1$, $\gamma_1 = 0.01$, $\Delta_2 = 0$, $W_{12} = -\Omega_2$, and $\theta = 15^{\circ}$.}
\vspace{-1em}
\end{figure}

We now compare the results of the present study with that of the three-level $V$-type atomic system considered in earlier publications \cite{wilson,autler}.
The four-level $Y$-type atom [see Fig. 1(a)] reduces to a three-level $V$-type atom if the ground atomic-state $|4\rangle$ is omitted. Figure 5 displays
the probe absorption and dispersion spectra of the $V$-type atomic system for the same probe and pump-fields interaction as shown in Fig. 1(b). On comparing
Figs. 2-4 with Fig. 5, it is seen that the dispersion spectra of the $V$ system takes a negative slope in the region between the gain peaks similar to that
of $Y$-type system. However, the negative dispersion is always accompanied with gain in the $V$ system and the gain at line centre $(\Delta_1 = 0)$ becomes
large as the separation of the gain peaks is reduced [compare dashed curves in Figs. 5(b) and 5(c)]. A distinguishing feature of the dispersion in the
$Y$-type atomic medium is that the anomalous dispersion occurs in the lossless (transparent) region between the gain peaks [see Figs. 2 and 3]. This region
of transparent dispersion with low absorption/gain will be preferred from an experimental point of view as the transparency condition minimizes reshaping of
the pulse in propagation
 \cite{wang}.

\vspace{-1em}
\section{origin of gain doublet}
\vspace{-1em}
In this section, we analyze the origin of gain features in the probe absorption spectrum. To this end, we first study the effects of the
decay-induced interference on the population distribution in the atomic levels. Since the probe field is assumed to be very weak, its effect on the
atomic population distribution can be considered negligible. Thus, we consider the equations (\ref{rho1})-(\ref{rho9}) with $\Omega_1=0$
and solve for the steady-state dynamics of the atom interacting with pump fields alone. The numerical results are obtained for the populations
as a function of the detuning $\Delta_2$ while keeping fixed the two-photon resonance condition $\Delta_2+\Delta_3=0$. In the absence of
interference term $(\gamma_{12}=0)$, there is zero population in the excited atomic state $|1\rangle$ and the atom behaves very much like a
three-level ladder system along the transitions $|2\rangle \leftrightarrow |3\rangle \leftrightarrow |4\rangle$ as expected. However, the
behavior of the atomic system is quite different when the interference term $(\gamma_{12}\neq0)$ is included as shown in Fig. 6.
The numerical results show that the steady-state population of the state $|1\rangle$ reaches a maximum value at $\Delta_2=\Delta_3=0$.
Also seen in Fig. 6 is an unexpected population inversion $(\bar{\rho}_{11}-\bar{\rho}_{33}\gg0)$ along the $|1\rangle\leftrightarrow|3\rangle$
transition at zero detunings of the pump fields. Thus, the origin of gain profiles in Fig. 2(b) can be attributed to the population
inversion along the probe transition.

To explore the reasons for the unexpected population inversion produced by decay-induced interference, we go to the dressed-atom description
of the atom-field interaction. In the dressed-state picture, the atomic states are obtained by diagonalizing the part of the Hamiltonian
(\ref{ham}) involving the interaction of the atom with pump fields only $(\Omega_1=0)$.  For simplicity, we consider only the case of zero
detunings of the fields, i.e., $\Delta_2=\Delta_3=0$. In this case, the Hamiltonian (\ref{ham}) has an eigenstate $|d\rangle$ with eigenvalue
zero:
\begin{equation}
|d\rangle = \frac{\Omega_3 |2\rangle - \Omega_2 |4\rangle}{\sqrt{\Omega_2^2 + \Omega_3^2}}. \label{dress1}
\end{equation}
In addition to the state $|d\rangle$, the atom has dressed states with non-zero eigenvalues which are defined as $H_I |\pm\rangle = \hbar
\lambda_{\pm} |\pm\rangle$. The dressed states $|\pm\rangle$ can be further expanded in terms of the bare atomic states as
\begin{equation}
|\pm\rangle = \frac{\Omega_2 |2\rangle - \lambda_{\pm} |3\rangle + \Omega_3 |4\rangle}{\sqrt{\Omega_2^2+\lambda_{\pm}^2+\Omega_3^2}},
\label{dress2}
\end{equation}
where $\lambda_{\pm} = \pm \sqrt{\Omega_2^2+\Omega_3^2}$. The states $|d\rangle$, $|\pm\rangle$ defined
in Eqs. (\ref{dress1}) and (\ref{dress2}) together with the excited atomic state $|1\rangle$ can be used as basis states for describing
the atomic dynamics.

\begin{figure}[t]
   \includegraphics[width=7cm]{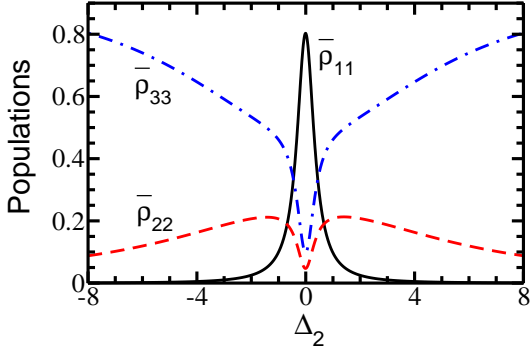}
   \vspace{-1em}
   \caption{(Color online) Steady-state populations $\bar{\rho}_{11}$(solid curve), $\bar{\rho}_{22}$(dashed curve), and $\bar{\rho}_{33}$(dot-dashed curve)
   as a function of the detuning $\Delta_2$ under the condition $\Delta_2+\Delta_3=0$. The parameters for the calculation are the
   same as in Fig. 2(b).}
   \vspace{-0.9em}
\end{figure}
A clear physical picture for the atomic population distribution can be obtained if the density matrix equation (\ref{mastereq}) is
written (with $\Omega_1=0$) in the dressed-state basis (\ref{dress1}) and (\ref{dress2}). We consider a special parameter choice
$\Omega_2=\Omega_3=\Omega$ and $W_{12}=-\sqrt{\Omega_2^2+\Omega_3^2}$ in which case the state $|1\rangle$ is degenerate with the
dressed state $|-\rangle$. In the high field limit $(\Omega_2,\Omega_3\gg\gamma_1,\gamma_2,\gamma_3)$, it is a good approximation to
neglect the coupling of density matrix elements associated with different frequencies in the dressed-state basis (secular approximation).
Under this approximation, the populations and coherences between dressed states obey equations of the form
\begin{flalign}
\dot{\rho}_{11} =&~\Gamma_{11}^{11} \rho_{11} + \Gamma_{++}^{11} \rho_{++} + \Gamma_{--}^{11} \rho_{--} + \Gamma_{dd}^{11} \rho_{dd}&
\nonumber \\
 & + \Gamma_{1-}^{11} \rho_{1-} + \Gamma_{-1}^{11} \rho_{-1},& \nonumber
\end{flalign}
\begin{flalign}
\dot{\rho}_{++} =&~\Gamma_{11}^{++} \rho_{11} + \Gamma_{++}^{++} \rho_{++} + \Gamma_{--}^{++} \rho_{--} + \Gamma_{dd}^{++} \rho_{dd}&
\nonumber \\
& + \Gamma_{1-}^{++} \rho_{1-} + \Gamma_{-1}^{++} \rho_{-1},& \nonumber
\end{flalign}
\begin{flalign}
\dot{\rho}_{--} =&~\Gamma_{11}^{--} \rho_{11} + \Gamma_{++}^{--} \rho_{++} + \Gamma_{--}^{--} \rho_{--} + \Gamma_{dd}^{--} \rho_{dd}&
\nonumber \\
& + \Gamma_{1-}^{--} \rho_{1-} + \Gamma_{-1}^{--} \rho_{-1}, & \label{rhod}
\end{flalign}
\begin{flalign}
\dot{\rho}_{dd} =&~\Gamma_{11}^{dd} \rho_{11} + \Gamma_{++}^{dd} \rho_{++} + \Gamma_{--}^{dd} \rho_{--} + \Gamma_{dd}^{dd} \rho_{dd}
\nonumber \\
& + \Gamma_{1-}^{dd} \rho_{1-} + \Gamma_{-1}^{dd} \rho_{-1},& \nonumber
\end{flalign}
\begin{flalign}
\dot{\rho}_{1-} =&~\Gamma_{11}^{1-} \rho_{11} + \Gamma_{++}^{1-} \rho_{++} + \Gamma_{--}^{1-} \rho_{--} + \Gamma_{dd}^{1-} \rho_{dd}&
\nonumber \\
& + \Gamma_{1-}^{1-} \rho_{1-} + \Gamma_{-1}^{1-} \rho_{-1},& \nonumber
\end{flalign}
\begin{figure}[t]
   \includegraphics[width=7cm]{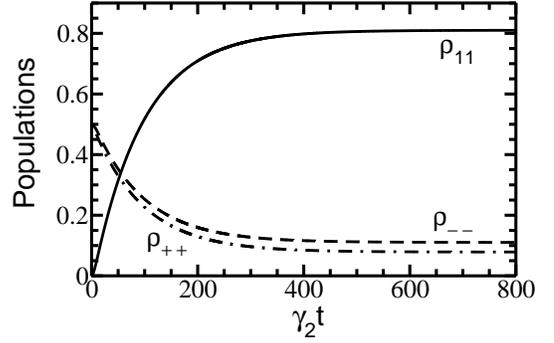}
   \vspace{-1em}
   \caption{Time evolution of atomic populations in the states $\rho_{11}$(solid curve), $\rho_{--}$(dashed curve), and $\rho_{++}$(dot-dashed curve).
   The parameters for the calculation are the same as in Fig. 2(b). The population $\rho_{dd}$ (not shown) is given by
   $\rho_{dd} = 1 -\rho_{11} - \rho_{++} - \rho_{--}$ due to the trace condition.}
   \vspace{-1em}
\end{figure}
where $\rho_{-1} = \rho_{1-}^{*} = \rho_{1-}$ is the (real) coherence between the states $|1\rangle$ and $|-\rangle$. The transition rates
($\Gamma$-terms) in Eqs. (\ref{rhod}) describe the decay as well as population transfers into each dressed state and its explicit
expressions are given in the Appendix. Note that the above equations differ from the usual rate equations for the
populations because of the coupling between the diagonal and off-diagonal elements. As seen from the expressions for the $\Gamma$
(see Appendix), the coherence $\rho_{1-}$ and the populations $\rho_{11}$ and $\rho_{--}$ are coupled only when the quantum
interference is present $(\gamma_{12}\neq0)$. It is thus expected that such coupling may lead to population trapping in the
degenerate atomic states $|1\rangle$ and $|-\rangle$ very similar to that reported earlier in degenerate $V$ systems \cite{gsa3}.
However, the dressed state $|-\rangle$ (containing the excited state $|2\rangle$) decays much faster compared to the state
$|1\rangle$ because of the assumption $\gamma_2 \gg \gamma_1$. Therefore, the population trapping will occur mainly in the
excited atomic state $|1\rangle$ due to decay induced interference. To confirm this, a numerical solution is obtained from
Eqs. (\ref{rhod}) for the time evolution of the population in states $|1\rangle$, $|+\rangle$, $|-\rangle$,
and $|d\rangle$ with the initial condition $\rho_{33}(0)=1$. In Fig. 7, we present the numerical results of the population
distribution for the same parameters of Fig. 2(b). It is seen from the graph that the population in the excited atomic state
$|1\rangle$ keeps increasing with time and reaches a maximum in the steady state consistent with the result for $\rho_{11}$ in
Fig. 6.
\vspace{-1em}
\section{effects of population trapping on the pump-field refraction index}
\vspace{-1em}
We now study the effects of the above population trapping on the pump-field line profiles. The dispersion/absorption of the
pump fields $E_2$ and $E_3$ are given by the real/imaginary parts of $\rho_{23}$ and $\rho_{34}$, respectively. It is well
known that a selective population of one of the dressed states will result in high refractive index with low absorption
for the driving field \cite{select1,select2}. In the previous studies, either adjusting the strength of the pumping field
\cite{select1} or the excitation by another field \cite{select2} was suggested to create a large population difference between
the dressed states. However, in the present case we show how decay-induced interference can be used to produce a large difference
of population in the dressed states. We consider the parameter choice $\Delta_2=\Delta_3=0$, $\Omega_2=\Omega_3=\Omega$,
$W_{12}=-\sqrt{\Omega_2^2+\Omega_3^2}$, and $\gamma_1\gg\gamma_2,\gamma_3$. As discussed in the previous section, population
trapping may occur in the degenerate states $|1\rangle$ and $|-\rangle$ due to decay-induced interference.
For $\gamma_1\gg\gamma_2$, the dressed state $|-\rangle$ decays slowly relative to the excited atomic state $|1\rangle$. Thus,
the population will get accumulated mainly in the state $|-\rangle$ resulting in a large population difference between the
$|-\rangle$ and $|+\rangle$ states.

In Fig. 8, we show the real and imaginary parts (solid curves) of the coherence $\rho_{23}$ as a function of the
detuning $\Delta_2$ in the presence $(\gamma_{12}\neq0)$ of interference term. For comparison, the result (dashed curves)
without quantum interference $(\gamma_{12}=0)$ is also shown. These numerical results are obtained by solving
Eqs. (\ref{rho1})-(\ref{rho9}) in steady state with $\Omega_1=0$. From the graphs, it is seen that the dispersion (real
part of $\rho_{23}$) has a peak at $\Delta_2=0$ and reaches the value of 0.3 for $\gamma_{12}= 0.98 \sqrt{\gamma_1 \gamma_2}$.
The effect of interference is found to reduce the absorption (imaginary part of $\rho_{23}$) at $\Delta_2=0$ in comparison with
that for $\gamma_{12}=0$. Similar results can be obtained for the coherence $\rho_{34}$ of the lower atomic transition. In the high
field limit $(\Omega_2,\Omega_3\gg\gamma_1,\gamma_2,\gamma_3)$, the coherences $\rho_{23}$ and $\rho_{34}$ can be worked out analytically
using the dressed-state basis. Under the assumption that coherences vanish between $|+\rangle$, $|-\rangle$ and $|d\rangle$ states, the
real parts of $\rho_{23}$ and $\rho_{34}$ are given by
\begin{figure}[t]
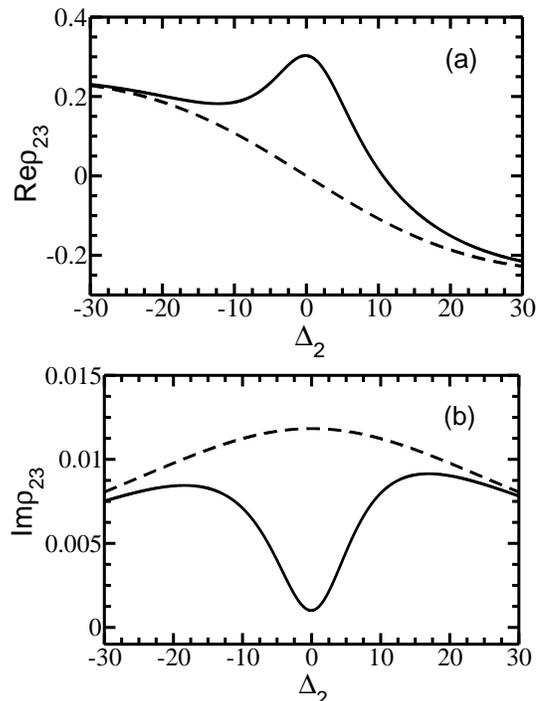

   \includegraphics[width=7cm]{fig8a.eps}
   \includegraphics[width=7cm]{fig8b.eps}
   \vspace{-1.5em}
   \caption{Real and imaginary parts of the coherence $\rho_{23}$ as a function of the pump-field detuning $\Delta_2$ under
   the condition $\Delta_2 + \Delta_3 = 0$. The parameters are  $\gamma_2 = 1$, $\gamma_1 = 5$,
   $\gamma_3 = 0.01$, $\Omega_2 = \Omega_3 = 30/\sqrt{2}$, and $W_{12} = -\sqrt{\Omega_2^2 +\Omega_3^2}$. The solid (dashed) curves
   are for $\theta = 10^{\circ}$ $(\theta = 90^{\circ})$. The results for $\rho_{34}$ (not shown) are identical to those of
   $\rho_{23}$.}
   \vspace{-1em}
\end{figure}
\begin{equation}
\hbox{Re}(\rho_{23}) = \hbox{Re}(\rho_{34}) \approx \frac{\rho_{--} - \rho_{++}}{2\sqrt{2}}. \label{cohere}
\end{equation}
Here, $\rho_{++}$ and $\rho_{--}$ refer to the steady-state population of the dressed states $|+\rangle$ and $|-\rangle$,
respectively. From Eq. (\ref{cohere}), it is clear that a large population difference $(\rho_{--}-\rho_{++}\gg 0)$
of the dressed states gives rise to a high refractive index (dispersion) for the pump fields. In the steady-state condition,
a long calculation for $\rho_{++}$ and $\rho_{--}$ using Eqs. (\ref{rhod}) leads to
\begin{flalign}
&\hbox{Re}(\rho_{23})=  \hbox{Re}(\rho_{34})&& \nonumber \\
&= \frac{\sqrt{2}\gamma_1}{[2\gamma_1(\gamma_3+2) (4\gamma_3+1)+(2\gamma_3+1) (2\gamma_3 (\gamma_3+2)+1)]}, &&
\label{cform}
\end{flalign}
where all the parameters have been scaled in units of $\gamma_2$. In deriving the result (\ref{cform}), we have assumed
the approximation $\gamma_{12} \approx \sqrt{\gamma_1\gamma_2}$. The formula (\ref{cform}) accounts well for the
dispersion (solid curve) shown in Fig. 8(a) at $\Delta_2=\Delta_3=0$. Note that the result (\ref{cform}) at zero detunings of
the pump fields is independent of $\Omega$ which we have confirmed from a direct numerical calculation of $\rho_{23}$
and $\rho_{34}$ from Eqs. (\ref{rho1})-(\ref{rho9}).
\vspace{-1em}
\section{conclusions}
\vspace{-1em}
In summary, we have investigated theoretically the effects of quantum interference in decay channels on the
pump-probe spectroscopy in Y-type atomic system. We find that a gain doublet appears in the absorption spectrum instead of the
usual absorption components. The origin of gain features has been explained as due to the population inversion
along the probe transition. The influence of quantum interference on the dispersion characteristics of the probe
field is also investigated. It is shown that the slope of the probe dispersion can be changed from positive to negative
values in the transparent region between the gain doublet. This allows, in principle, the probe pulse propagation to be switched
from subluminal to superluminal domain. Further, we have also shown the possibility of refractive-index enhancement for
the pump fields via decay-induced interference.

It should be noted that the results predicted in this work are based on interference effects in spontaneous emission which require
the existence of non-orthogonal dipole moments of the atomic transitions. In practice, it is difficult to meet this requirement in
real atoms. Several alternative schemes involving coherent- and dc-field-induced splitting of atomic levels \cite{patnaik1,dcfield},
cavities with preselected polarisation \cite{patnaik2}, and decay in anisotropic vacuum \cite{gsa5} have been proposed
to realize interference effects surpassing the requirement of non-orthogonal dipole transitions. For a realistic example of the
$Y$-type configuration of energy levels, one can choose the hyperfine states of the rubidium atom. The hyperfine states
$|5S_{1/2},F=3\rangle$ and $|5P_{3/2},F^{'}=4\rangle$ in the $^{85}\hbox{Rb}$ atom may be chosen as the states $|4\rangle$ and 
$|3\rangle$, respectively, whereas the states $|5D_{5/2},F^{''}=3\rangle$ and $|5D_{5/2},F^{''}=4\rangle$ in the atom could be the  
excited states $|1\rangle$ and $|2\rangle$ with small energy separation \cite{song}. Finally, we note that doppler broadening and 
collisional dephasing are other dominant mechanisms to suppress interference effects in spontaneous decay \cite{dopler}. However, 
these motional effects of the atoms can be avoided by using cold atomic ensemble as in the experiment of Hau {\it et al.} \cite{hau}. 
\vspace{-1em}
\section*{ACKNOWLEDGMENT}   
\vspace{-1em}
The author thanks Hebrew Benhur Crispin for helpful discussions.
\vspace{-1em}
\appendix*
\section{Decay and transition rates of the dressed states}
\vspace{-1em}
The $\Gamma$ of Eqs. (\ref{rhod}) are given by
\begin{flalign}
&\Gamma_{11}^{11} = -2 \gamma_1,~~~~\Gamma_{++}^{11} = \Gamma_{--}^{11} = \Gamma_{dd}^{11} = 0,& \nonumber \\
& \Gamma_{1-}^{11} = \Gamma_{-1}^{11} = -\gamma_{12}/2,&
\end{flalign}
\vspace{-2em}
\begin{flalign}
&\Gamma_{11}^{++} = \gamma_1,~~~~\Gamma_{++}^{++} = -(\gamma_2 + 3 \gamma_3)/4,~~~~\Gamma_{dd}^{++} = \gamma_2/2,
&\nonumber \\
&\Gamma_{--}^{++} = (\gamma_2 + \gamma_3)/4,~~~~\Gamma_{1-}^{++} = \Gamma_{-1}^{++} = \gamma_{12}/2,&
\end{flalign}
\vspace{-2em}
\begin{flalign}
&\Gamma_{11}^{--} = \gamma_1,~~~~\Gamma_{--}^{--} = -(\gamma_2 + 3 \gamma_3)/4,~~~~\Gamma_{dd}^{--}=\gamma_2/2,&
\nonumber \\
&\Gamma_{++}^{--} = (\gamma_2 + \gamma_3)/4,~~~~\Gamma_{1-}^{--} =\Gamma_{-1}^{--} = 0,&
\end{flalign}
\vspace{-2em}
\begin{flalign}
&\Gamma_{11}^{dd} = \Gamma_{1-}^{dd} = \Gamma_{-1}^{dd} = 0,~~~~\Gamma_{dd}^{dd} = -\gamma_2, &\nonumber \\
&\Gamma_{++}^{dd} = \Gamma_{--}^{dd} = \gamma_3/2,&
\end{flalign}
\vspace{-2em}
\begin{flalign}
&\Gamma_{11}^{1-} = \Gamma_{--}^{1-} = -\gamma_{12}/2,~~~~\Gamma_{++}^{1-} = \Gamma_{dd}^{1-} =
\Gamma_{-1}^{1-} = 0,& \nonumber \\
&\Gamma_{1-}^{1-} = -(4 \gamma_1 + \gamma_2 + 2 \gamma_3)/4.&
\end{flalign}

\end{document}